\documentclass[aps,11pt,nofootinbib]{revtex4-1}
\pdfoutput=1
\usepackage{graphicx}
\usepackage[margin=1in]{geometry}
\usepackage{amsmath}
\usepackage{amssymb}
\usepackage{color}
\usepackage[utf8]{inputenc}

\makeatletter
\renewcommand\@make@capt@title[2]{%
  \@ifx@empty\float@link{\@firstofone}{\expandafter\href\expandafter{\float@link}}%
   {\textbf{#1}}\@caption@fignum@sep#2\quad
}%
\makeatother

\newcommand{\fnl}{f_{\rm NL}}

\begin{document}

\author{Sam Young$^1$}
\email{syoung@mpa-garching.mpg.de}
\author{Christian T. Byrnes$^{2}$}
\email{C.Byrnes@sussex.ac.uk}

\affiliation{\\ 1) Max Planck Institute for Astrophysics, Karl-Schwarzschild-Strasse 1, 85748 Garching bei Muenchen, Germany} 

\affiliation{\\ 2) Department of Physics and Astronomy, University of Sussex, Brighton BN1 9QH, United Kingdom\\ }

\date{\today}

\title{Initial clustering and the primordial black hole merger rate}

\begin{abstract}


If the primordial curvature perturbation followed a Gaussian distribution, primordial black holes (PBHs) will be Poisson distributed with no additional clustering. We consider local non-Gaussianity and its impact on the initial PBH clustering and mass function due to mode coupling between long and short wavelength modes. We show that even a small amount of non-Gaussianity results in a significant enhancement on the PBH initial clustering and subsequent merger rate and that the PBH mass function shifts to higher mass PBHs. However, as the clustering becomes strong, the local number density of PBHs becomes large, leading to a large theoretical uncertainty in the merger rate.

\end{abstract}

\maketitle

\tableofcontents

\section{Introduction}

Primordial black holes (PBHs) are black holes which may have formed very early in the history of the universe. There are several mechanisms by which they may have formed (for example, from cosmic strings \cite{Hawking:1987bn} or bubble collisions \cite{Hawking:1982ga}), but we will here focus on PBHs which form from the collapse of large density perturbations, as proposed in \cite{Hawking:1971ei,Carr:1974nx}. If a density perturbation has a large enough amplitude, it will collapse to form a PBH upon horizon entry. Such perturbations are sourced during cosmological inflation, where quantum fluctuations can become classical density perturbations as they exit the horizon, which then go on to re-enter the horizon following the end of inflation, and are also responsible for the growth of cosmological structure. The amplitude of such perturbations on small scales that form PBHs is required to be orders of magnitude larger than that observed on cosmological scales, although there are many models which do make this prediction (for example, \cite{Drees:2011hb,Bugaev:2013fya,Ozsoy:2018flq,GarciaBellido:1996qt,Lyth:2012yp,Bugaev:2011wy,Ballesteros:2018wlw}, amongst many others).

The typical assumption is that the perturbations from which PBHs form have a Gaussian distribution, although recent papers have discussed the fact that, even if the curvature perturbation $\zeta$ is Gaussian, the density perturbations $\delta$ are not \cite{Young:2019yug,Yoo:2018esr,Kawasaki:2019mbl,Kalaja:2019uju} - and it is the density which one should consider when investigating PBH formation \cite{Young:2014ana}. 
There has been extensive research over the last decades to study the effects of primordial non-Gaussianity (PNG) on the abundance of PBHs \cite{Bullock:1996at,Ivanov:1997ia,Byrnes:2012yx,Shandera:2012ke,Young:2013oia,Young:2015cyn,Franciolini:2018vbk,Yoo:2019pma,Atal:2019cdz,Atal:2018neu} - finding that the presence of PNG can have contrasting effects on the abundance depending on the type of non-Gaussianity considered, but we note that local non-Gaussianity normally greatly enhances the abundance of PBHs, even with a negative skewness ($\fnl<0$) unless the power spectrum is narrowly peaked \cite{Young:2014oea}.  References \cite{Tada:2015noa,Young:2015kda,Suyama:2019cst,Matsubara:2019qzv} also studied the clustering of PBHs in non-Gaussian conditions, finding that even a small amount of PNG leads to signficant clustering, which is otherwise absent. 

In this paper, we will go beyond the work of previous papers, which have mostly focussed on the effect of PNG on the abundance of PBHs, and consider other key potential observables of PBHs as a function of the level of PNG: the PBH merger rate today and the PBH mass function. We will focus on a model where PNG leads to a strong coupling between modes of different scales, which leads to a spatially dependent PBH formation rate. different formation rates of PBHs in different regions of the universe.

Recent interest has often focused on the LIGO mass range, because this range is currently testable and also because the low effective spin of the observed merger events could hint at a primordial origin \cite{Clesse:2017bsw,Mirbabayi:2019uph,Postnov:2019tmw,Fernandez:2019kyb,He:2019cdb}. Since the first direct detection of a BH merger, two papers \cite{Bird:2016dcv,Clesse:2016vqa} claimed that the PBH merger rate would match that detected by LIGO if all of the DM consisted of PBHs, but these papers calculated the ``late'' time merger rate caused by PBHs which form binary pairs in the late universe. However, the merger rate of PBH binaries which form in the early universe, before matter-radiation equality, is expected to dominate the merger rate today \cite{Nakamura:1997sm,Ioka:1998nz,Sasaki:2016jop,Ali-Haimoud:2017rtz,Chen:2018czv,Raidal:2017mfl,Ballesteros:2018swv,Raidal:2018bbj} which leads to the tightest observational constraint on the allowed fraction of PBHs in this mass range, with the constraint being somewhere between 1--10$\%$ \cite{Vaskonen:2019jpv}. Those papers assumed that the PBHs were not initially clustered, but \cite{Bringmann:2018mxj} showed that strong initial clustering actually tightened the constraints. However, they applied the standard formalism used to calculate the merger rate even in regions where the PBH density was extremely large, which was subsequently shown to be a poor approximation because PBH binaries are often disrupted in regions with a large density of PBHs \cite{Raidal:2018bbj}.

The paper is organised as follows: in section \ref{sec:PBHbackground} we will discuss the formation and abundance of PBHs in the absence of PNG, and develop this to account for PNG in section \ref{sec:NG}. In section \ref{sec:mass} we will derive an expression for the mass function of PBHs dependent on the statistics of the early universe, for PBHs forming at a single mass scale. In section \ref{sec:powerSpectrum}, we derive an expression relating the power spectrum of perturbations to the PBH abundance, and finally provide a summary of the key findings of the paper in section \ref{sec:summary}.

\section{Primordial black hole abundance in a Gaussian universe}
\label{sec:PBHbackground}

Density perturbations above a certain threshold value will collapse upon horizon re-entry. The most robust criterion to use is the volume-averaged density contrast (also called the smoothed density contrast) \cite{Young:2019osy}. Smoothing with a real-space top-hat function, the threshold value can take a value in the range $0.44 \leq \delta_c \leq 0.67$, depending on the shape of the perturbation \cite{Musco:2018rwt}. For a typical profile shape  expected from the primordial perturbations generated, this takes a value $\delta_c=0.51$ \cite{Germani:2018jgr,Young:2019osy}, which will be used throughout this paper. The effect of non-Gaussianity in the primordial distribution of matter and its effect on the profile shape, and its corresponding effect on the threshold value, was studied in a recent paper \cite{Kehagias:2019eil} which found that the critical value of the volume-averaged density contrast does not change significantly.

Perturbations which form PBHs in the early universe are necessarily rare. For example, in the case of solar mass PBHs, only one billionth of the universe needs to collapse into a PBH in order for PBHs to make up all of the dark matter today. Rare perturbations are expected to be approximately spherically symmetric \cite{Bardeen:1985tr}. Assuming spherical symmetry therefore, the smoothed density contrast $\delta_R$ can be related to the curvature perturbation $\zeta$ in the comoving synchronous gauge as
\begin{equation}
\delta_R = - \frac{2}{3} R \zeta^\prime(R) \left( 2 + R \zeta^\prime(R)\right),
\label{eqn:NLdensity}
\end{equation}
where $R$ is the smoothing scale ($r_m$ is typically used in the literature to represent the characteristic scale of a perturbation in the calculation of $\delta_c$, also representing the correct smoothing scale for a given perturbation), and the prime denotes a derivative with respect to the radial co-ordinate $r$. This equation allows us to relate the statistics of $\delta_R$, which determines PBH formation \cite{Young:2014ana}, to the statistics of $\zeta$. The linear component of $\delta_R$, (which we will label as $\delta_1$), is given by
\begin{equation}
\delta_1 = -\frac{4}{3}R \zeta^\prime(R).
\end{equation} 
The mass of a PBH that forms from the perturbations depends upon the scale and amplitude of the perturbation as  
\begin{equation}
M_{PBH} = \mathcal{K} M_h \left( \delta_R - \delta_c \right)^\gamma,
\label{eqn:scalingLaw}
\end{equation}
where $\mathcal{K}\approx 4$ for most profile shapes, and $\gamma = 0.36$ during radiation domination \cite{Musco:2004ak,Musco:2008hv,Musco:2012au,Young:2019yug,Escriva:2019nsa}, and $M_h$ is the horizon mass of a flat FRW universe at the time the Hubble horizon is the same scale as the perturbation.

We will begin by assuming that $\zeta$ has a Gaussian distribution, before accounting for the effect of primordial non-Gaussiniaty in section \ref{sec:NG}. Following the method described in \cite{Young:2019yug}, the mass fraction of the universe collapsing to form PBHs at the time of formation is given by
\begin{equation}
\beta = \int\limits_{\delta_{c,1}}^{\frac{4}{3}} \mathrm{d}\delta_1 \frac{\mathcal{K}}{4\pi^2} \left( \delta_1-\frac{3}{8}\delta_1^2 - \delta_c \right)^\gamma  \left( \frac{\sigma_1}{\sigma_0} \right)^3 \left( \frac{\delta_1}{\sigma_0} \right)^3 \exp \left( -\frac{\delta_1^2}{2\sigma_0^2} \right),
\label{eqn:beta}
\end{equation}
where $\delta_{c,1}$ is the critical value for the linear component,
\begin{equation}
\delta_{c,1} = \frac{4}{3}\left( 1-\sqrt{1-\frac{3}{2}\delta_c}\right),
\end{equation}
and finally $\sigma_j$ are moments of the power spectrum $\mathcal{P}_{\delta_1}$, given by
\begin{equation}
\sigma_j^2 = \int\limits_0^\infty \frac{\mathrm{d}k}{k} \mathcal{P}_{\delta_1}(k, \eta) \left( \frac{k}{aH} \right)^{2j},
\end{equation}
where $\eta$ is conformal time. During radiation domination, the power spectrum $\mathcal{P}_{\delta_1}$ can be calculated from the curvature perturbation power spectrum $\mathcal{P}_\zeta$ as
\begin{equation}
\mathcal{P}_{\delta_1}(k,\eta) =  \frac{16}{81}\int\limits_0^\infty \frac{\mathrm{d}k}{k}\left(\frac{k}{aH}\right)^4 \tilde{W}^2(k, R) T^2 (k, \eta) \mathcal{P}_\zeta(k),
\end{equation}
where $\tilde{W}(k, R)$ is the Fourier transform of the (real-space top-hat) smoothing function with a smoothing scale $R=(aH)^{-1}$, which is the Hubble scale,
\begin{equation}
\tilde{W}(k,R)= 3 \frac{\sin(kR)-kR\cos(kR)}{(k R)^3},
\end{equation}
and $T(k,\eta)$ is the linear transfer function at a time $\eta$,
\begin{equation}
T(k,\eta)= 3 \frac{\sin(k\eta/\sqrt{3})-(k\eta/\sqrt{3})\cos(k\eta/\sqrt{3})}{(k\eta/\sqrt{3})^3}.
\end{equation}
Importantly, on super-horizon scales, $\mathcal{P}_{\delta_1}$ is proportional to the wave number to the fourth power, $\mathcal{P}_{\delta_1}\propto k^{4}$. We therefore conclude that (when perturbations follow a Gaussian distribution), super-horizon modes at the time of formation have a negligible effect on PBH formation.

We will now discuss, briefly, the consequences of this in terms of the primordial clustering of PBHs arising from scale-independent bias. Scale-independent bias essentially happens because small-scale peaks can be situated inside large-scale peaks - meaning that many more small-scale peaks are likely to be above the threshold value in regions of the universe where there already exists a large-scale peaks, leading to the conclusion that PBHs preferentially form in large-scale overdense regions. However, because such large-scale overdensities are super-horizon at the time of PBH formation, they are negligibly small - and thus the effect of scale-independent bias is negligible on scales significantly larger than the scale of the PBH. This has been well documented in numerous papers \cite{Tada:2015noa,Young:2015kda,Ali-Haimoud:2018dau,Ballesteros:2018swv,MoradinezhadDizgah:2019wjf,Suyama:2019cst}, and also means that the spatial distribution of PBHs is expected to be Poissonian in the case of Gaussian statistics.

As described above, the perturbations which form PBHs are necessarily rare in order that their abundance does not quickly come to dominate the energy of the universe. For PBHs of around 1 solar mass, an initial abundance of $\beta \sim 10^{-9}$ will cause them to dominate the energy fraction of the universe after the time of matter-radiation equality. Since the size of a PBH is approximately equal to the scale of the perturbation it formed from at horizon re-entry, the average separation between PBHs will be $\mathcal{O}(10^3)$ times greater than the horizon scale at the time of formation. If we take this scale as being the smallest scale relevant for the clustering of PBHs, then the amplitude of modes which may give rise to scale-independent bias is suppressed by a factor of at least $\mathcal{O}(10^{-6})$ - and may safely be neglected.

\section{Primordial black hole abundance in a non-Gaussian universe}
\label{sec:NG}

We will now discuss the effects of non-Gaussianity on the abundance and initial clustering of PBHs. We will study the effect of local-type non-Gaussianity to second order, where the curvature perturbation $\zeta$ is related to the Gaussian-distributed $\zeta_G$ as \cite{Komatsu:2001rj}
\begin{equation}
\zeta = \zeta_G+\frac{3}{5} f_{\mathrm{NL}}\left( \zeta_G^2 - \langle\delta_G^2\rangle \right),
\label{eqn:localNG}
\end{equation}
where $f_{\mathrm{NL}}$ is the non-linearity parameter (the $\langle\delta_G^2\rangle$ term ensures that the mean of $\zeta$ remains zero while $\zeta_G$ has a mean of zero). This model for non-Gaussianity is useful for several reasons: firstly, it allows us to make an analytic estimate of the effects of non-Gaussianity\footnote{Other bispectrum shapes were considered numerically in \cite{Young:2015cyn}, and found to have a qualitatively similar effect on the abundance of PBHs.}, and secondly, local-type non-Gaussianity contains a strong coupling between the small-scale modes on which PBHs form, and the large-scale modes at which they cluster. It is also the typical type of non-Gaussianity generated during multiple-field inflation, e.g.~\cite{Byrnes:2014pja} for a review.

It has been shown that local-type non-Gaussianity can lead to large dark-matter isocurvature modes, which are tightly constrained on CMB scales \cite{Young:2015kda,Tada:2015noa}, leading to strong constraints on the non-Gaussianity in the scenario that a significant fraction of dark matter is composed of PBHs, $f_{\mathrm{NL}} < \mathcal{O}(10^{-3})$. However, this constraint only applies to mode-coupling to scales large enough to be observable by the Planck satellite on the CMB and also does not apply in the case of single-field inflation. The constraints can be avoided if the bispectrum is assumed to be negligibly small on CMB scales, or uncorrelated to the PBH forming scales, but it may be larger on intermediate scales - which, as we will see, will result in significant clustering of PBHs on such scales. 

\begin{figure}
 \centering
 \includegraphics[width=\textwidth]{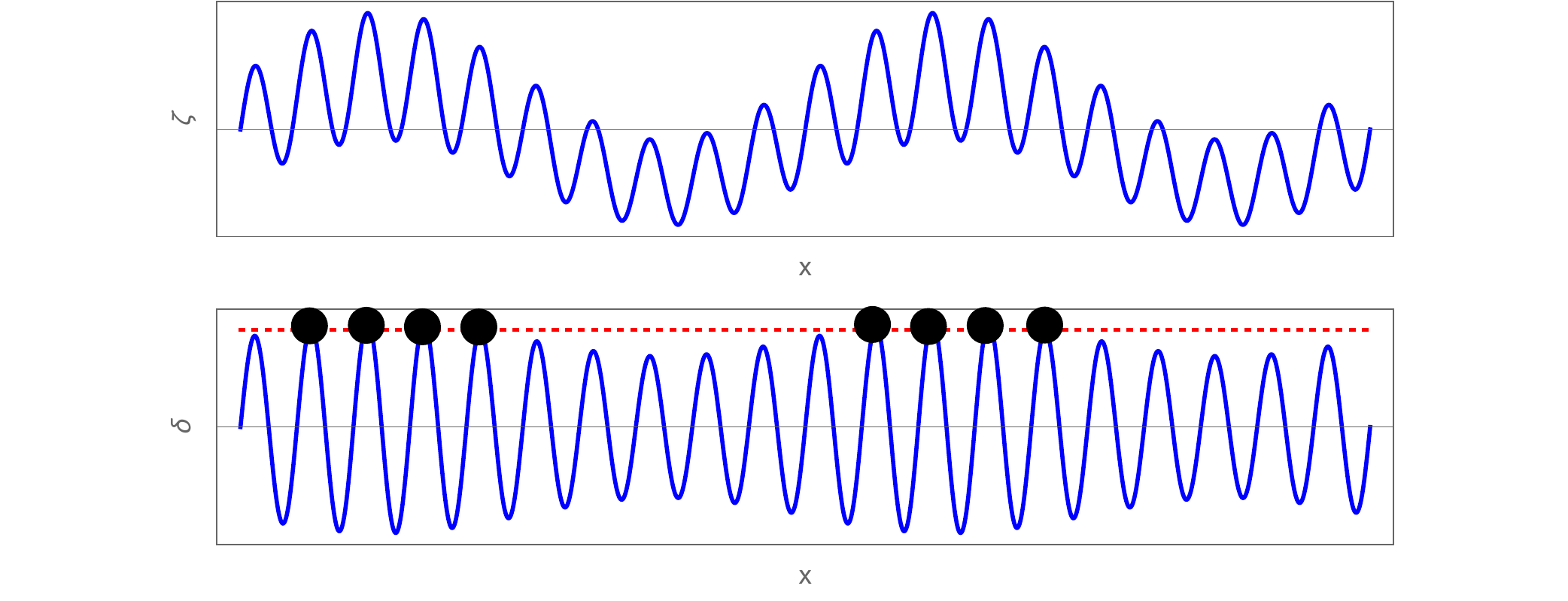} 
 \caption{ A schematic plot of a universe containing exactly 2 modes. The x-axis represents spatial coordinates, whilst the y-axis represents the curvature perturbation $\zeta$ and the density contrast $\delta$ in the top and bottom plots respectively, at the time the small-scale wavelength enters the horizon. The red dashed line represents the threshold value in the density contrast in order for a PBH to form, whilst the black circles represent locations where a PBH will form. In a non-Gaussian, the two modes couple to each other, and the amplitude of the short-wavelength mode depends on the amplitude of the long-wavelength mode. This leads to larger perturbations in some regions of the universe - and consequently enhanced PBH formation in those regions.}
\label{fig:modalCoupling}
\end{figure}

The volume-averaged density contrast $\delta_R$ is related to the curvature perturbation $\zeta$ during radiation domination by equation \eqref{eqn:NLdensity}. Differentiating equation \eqref{eqn:localNG} to find $\zeta'$ gives
\begin{equation}
\zeta' = \left(1+\frac{6}{5} f_{\mathrm{NL}}\zeta_G \right) \zeta_G^\prime,
\label{eqn:zetaPrime}
\end{equation}
which gives an expression for $\delta_R$ in terms of the Gaussian variable $\zeta_G$,
\begin{equation}
\delta_R = - \frac{2}{3} r_m \left(1+\frac{6}{5} f_{\mathrm{NL}}\zeta_G(r_m) \right) \zeta_G^\prime(r_m) \left( 2 + r_m \left(1+\frac{6}{5} f_{\mathrm{NL}}\zeta_G(r_m) \right) \zeta_G^\prime(r_m)\right).
\label{eqn:PNGdelta}
\end{equation}
This expression now depends not only upon derivatives of $\zeta$, but on the absolute value of $\zeta$ itself. Therefore, the argument made in the previous section that large-scale modes do not affect PBH formation (and that therefore PBHs do not initially form in clusters) is no longer valid. This dependance of $\delta$ on the absolute value of $\zeta$ leads to modal coupling and initial clustering of PBHs, an example is shown schematically in figure \ref{fig:modalCoupling}. The consequences of this were discussed in detail in \cite{Young:2014oea,Young:2015kda,Tada:2015noa}, which discussed how the abundance of PBHs and the amplitude of of isocurvature modes is affected. 

Over the course of this paper, we will extend the calculation to account for the non-linear relationship between $\zeta$ and $\delta$ and the critical scaling relationship, equation \eqref{eqn:scalingLaw}, as well as going on to calculate the effect on the mass function of PBHs, the primordial clustering, and the observed merger rate today.

In principle, $\zeta_G$ and its derivative $\zeta_G'$ will be correlated. For example, a short distance away from the centre of a peak in $\zeta_G$, $\zeta_G$  is likely to be positive, whilst $\zeta_G'$ is likely to be negative - and the opposite is true for troughs. However, the exact relation between $\zeta_G$ and $\zeta_G^\prime$ depends on the profile shapes of the perturbation, which itself depends on the mechanism by which the perturbations were initially generated, and such a consideration goes beyond the scope of this work
\footnote{However, the ``typical'' profile shape of perturbations from non-Gaussian initial conditions and its effect on the formation threshold have recently been considered \cite{Atal:2019cdz, Kehagias:2019eil}.}.
We will therefore make the assumption that $\zeta_G$ is independent of $\zeta_G^\prime$, which can be treated as equivalent to assuming that the distribution of $\zeta$ is Gaussian in small regions of the universe, with the variance of perturbations varying from region to region. To validate this statement, we will split $\zeta_G$ into long- and short-wavelength components,
\begin{equation}
\zeta_G = \zeta_s + \zeta_l,
\end{equation}
where only the short-wavelength modes will be relevant for PBH formation (the longer-wavelength modes being super-horizon at the time of formation). Due to the Gaussianity of $\zeta_G$, $\zeta_s$ will be uncorrelated with $\zeta_l$. Inserting this into equation \eqref{eqn:localNG} and taking the first derivative (as is necessary to calculate the formation criterion given by equation \eqref{eqn:NLdensity}) gives
\begin{equation}
\zeta^\prime = \left( 1 + \frac{6}{5} f_{\mathrm{NL}} \left( \zeta_s+\zeta_l \right) \right)\left( \zeta_s^\prime+\zeta_l^\prime \right).
\end{equation}
The derivatives of $\zeta_l$ will be negligibly small compared to derivatives of $\zeta_s$, and so will be neglected. Finally, if we make the approximation that $\zeta_s$ has a Gaussian distribution and therefore set $f_{\mathrm{NL}}\zeta_s = 0$, we obtain the relation
\begin{equation}
\zeta^\prime = \left(1+\frac{6}{5} f_{\mathrm{NL}}\zeta_l \right) \zeta_s^\prime,
\end{equation}
which matches equation \eqref{eqn:zetaPrime}, with $\zeta_G \rightarrow \zeta_l$ and $\zeta_G' \rightarrow \zeta_s'$, where the zeroth and first-order derivatives uncorrelated. For the sake of clarity in the rest of this paper, we will use the notation $\zeta_l$ and $\zeta_s'$.

We note that the assumption of Gaussianity on the PBH forming scales will not significantly affect the main results of this paper
\footnote{We note that this will not be true in the case of $f_{\mathrm{NL}}\lesssim-1$. In this case, very small changes in the amplitude of the power spectrum can have an overwhelmingly large impact on the local PBH number density.}: 
firstly, whilst PBH abundance depends on the level of non-Gaussianity, it is degenerate with the amplitude of the small-scale power spectrum, which is treated here as a free parameter. Secondly, whilst the PBH merger rate does depend on the mass function of PBHs which itself will depend on the small-scale non-Gaussianity, this is a relatively small effect compared to the effect of the PBH abundance. 

In a given region of the universe, with a constant $\zeta_l$, we can then simply treat $\zeta^\prime$ as a Gaussian variable, where the amplitude of the perturbation is simply modified by a factor $(1 + \frac{6}{5} f_{\mathrm{NL}}\zeta_l)$, which is essentially modifying the local power spectrum. Writing the variance of the small-scale perturbations $\sigma_s^2$ as a function of $f_{\mathrm{NL}}\zeta_l$ and the ``background'' (i.e.~long wavelength) variance $\sigma_b^2$ gives
\begin{equation}
\sigma_s^2 = \left(1 + \frac{6}{5} f_{\mathrm{NL}}\zeta_l\right)^2 \sigma_b^2.
\label{eqn:sigmaBias}
\end{equation}
As described in section \ref{sec:PBHbackground}, the abundance of PBHs depends exponentially upon $\sigma^2$, and so even small changes in $\sigma^2$ can mean a large change in the PBH abundance: it will be greatly amplified in regions of positive $f_{\mathrm{NL}}\zeta_l$, and greatly reduced for negative values - an effect often described as scale-dependant bias.

In small regions of the universe, where $\zeta_s^\prime$ can be treated as Gaussian, we follow the method of \cite{Young:2019yug}, and the local abundance of PBHs at the time of formation can be calculated with equation \eqref{eqn:beta} as a function of $f_{\mathrm{NL}}\zeta_l$ and  $\sigma_b^2$ by substituting $\sigma_0\rightarrow\sigma_s$, the notation $\beta_{\rm local}$ will be used to describe the local value of $\beta$ in a small patch of the universe with constant $\zeta_l$. The fraction $\sigma_1/\sigma_0$ appearing in equation \eqref{eqn:beta} is not affected by the long wavelength mode because $\sigma_1$ and $\sigma_0$ are changed by the same factor.

For simplicity, we will here assume that the power spectrum has a Dirac-delta peak at some scale $k_*$, and takes some smaller (but not necessarily constant) value at all other scales,
\begin{equation}
\mathcal{P}_\zeta=\mathcal{A}_s \delta_D(\mathrm{ln}(k/k_*))+\mathcal{P}_b(k), 
\label{eqn:diracPower}
\end{equation}
such that all PBH formation occurs at a single scale. The term $\mathcal{P}_b(k)$ represents the (smaller) value of the power spectrum at all other scales, $k\neq k_*$. We have made this assumption for simplicity, although our method can be extended to account for any power spectrum, which may have an extended PBH formation time.

The parameter $\delta_\beta$ is introduced to describe the relative change in the abundance of PBHs,
\begin{equation}
\delta_{PBH} = \frac{\beta_{\rm local}(\sigma_s^2)-\beta_0(\sigma_b^2)}{\beta_0(\sigma_b^2)},
\label{eqn:deltaPBH}
\end{equation}
where $\beta_0(\sigma_b^2)$ simply gives the ``background'' value for $\beta$ in the absence of a large-scale $\zeta$ perturbation\footnote{However, note that using this definition means that the background $\beta_0$ will not be the mean value for $\beta$, unless $\langle \left( f_{\mathrm{NL}}\zeta_l\right)^2\rangle$ is small, but does give a convenient definition that doesn't depend on the form of the power spectrum, or total abundance of PBHs.}.

\begin{figure*}[t!]
 \centering
  \includegraphics[width=0.5\textwidth]{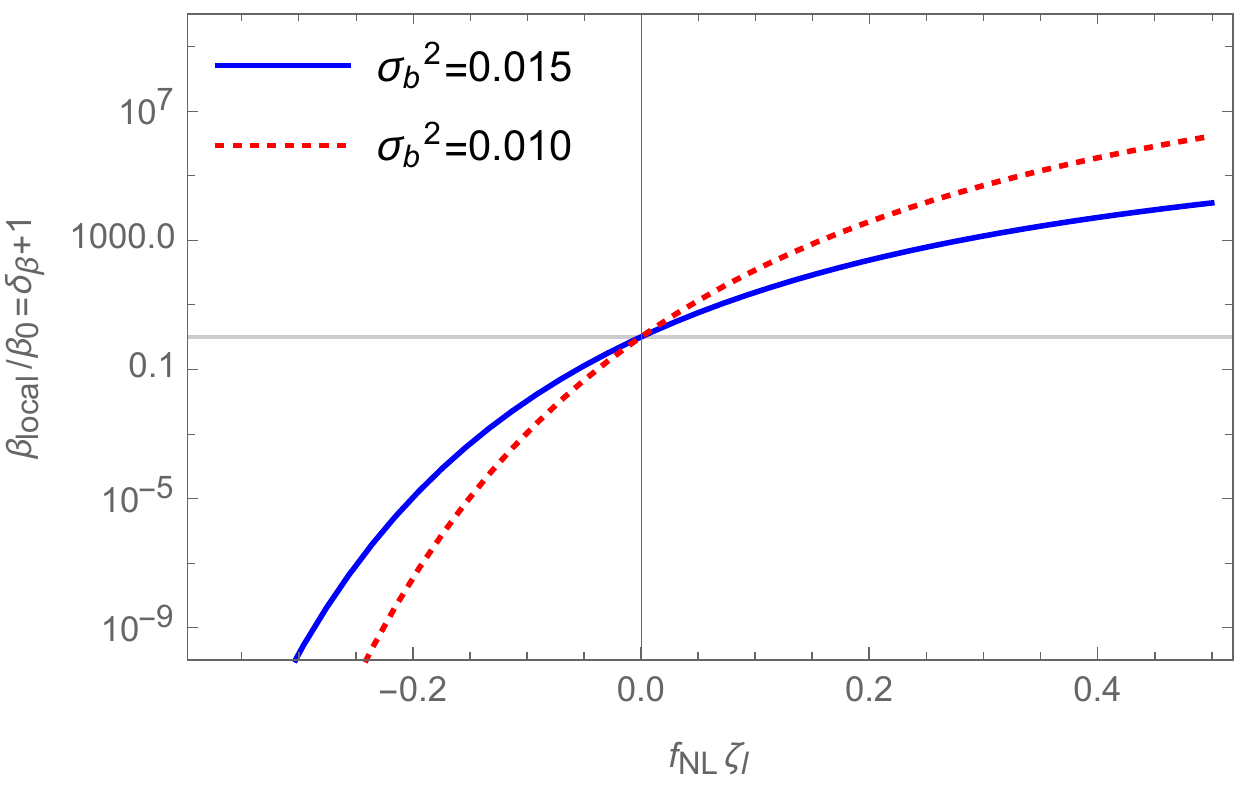} 
  \includegraphics[width=0.48\textwidth]{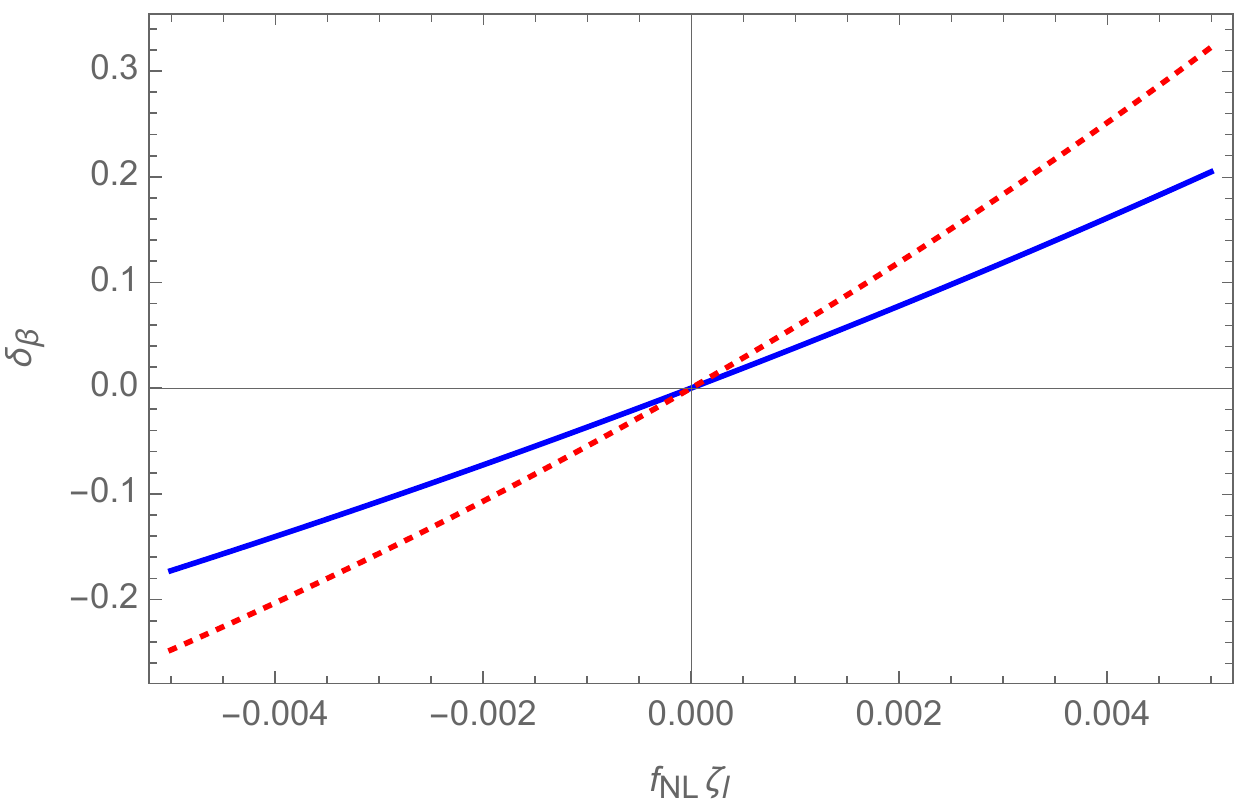} 
  \caption{ Perturbations to the initial abundance of PBHs are shown as a function of the non-Gaussianity parameter $f_{\mathrm{NL}}$ and a super-horizon curvature perturbation $\zeta_l$. The background value for the small-scale variance has been chosen as $\sigma_b^2=0.015$ and $0.010$ for the solid blue and red dotted lines respectively. The left plot has a logarithmic scale, and $\delta_\beta+1$ is therefore plotted. The right plot shows a zoomed in plot for small values of $f_{\mathrm{NL}}\zeta_l$.} 
  \label{fig:deltabeta}
\end{figure*}

Figure \ref{fig:deltabeta} shows $\delta_\beta$ as a function of $f_{\mathrm{NL}}\zeta_l$, where $\sigma_b$ has been selected such that PBHs compose 1\% and 0.1\% of dark matter for the solid blue and dotted red lines respectively. The right-hand plot shows a zoom of the central region. It can be immediately seen that $f_{\mathrm{NL}}\zeta_l$ has an impact on the abundance of PBHs orders of magnitude larger than $f_{\mathrm{NL}}\zeta_l$, and a perturbative treatment will not provide accurate results except for very small values of $f_{\mathrm{NL}}\zeta_l$
\footnote{Reference \cite{Young:2015kda} studied such small values on CMB scales, and derived a linear expression for the bias factor for the scale-dependent bias arising from the non-Gaussianity parameters $f_{\rm NL}$ and $g_{\rm NL}$. 
References \cite{Young:2015kda,Tada:2015noa}  studied perturbations on CMB scales, where $\zeta=\mathcal{O}(10^{-5})$, and found $f_{\mathrm{NL}}<\mathcal{O}(10^{-3})$ in order that observational bounds on isocurvature modes were not exceeded.}.

The total abundance of PBHs at formation may be obtained by integrating the local values of the abundance over the entire range of values of  $f_{\mathrm{NL}}\zeta_l$,
\begin{equation}
\beta\left( \sigma_b^2, \langle f_{\mathrm{NL}}^2\zeta_l^2 \rangle \right)=\frac{1}{\sqrt{2 \pi  \langle f_{\mathrm{NL}}^2\zeta_l^2 \rangle}}\int\limits_{-\infty}^\infty \mathrm{d}(f_{\mathrm{NL}}\zeta_l) \beta_{\rm local} \left( \left(1+\frac{6}{5}f_{\mathrm{NL}}\zeta_l\right)^2\sigma_b^2 \right)\exp\left( -\frac{ f_{\mathrm{NL}}^2\zeta_l^2}{2\langle f_{\mathrm{NL}}^2\zeta_l^2 \rangle} \right).
\end{equation}
It is found that the modal coupling due to non-Gaussianity always increases the number of PBHs form (see \cite{Young:2014oea}, where this was studied extensively). 

In the following sections, we will discuss the implications of the scale-dependent bias on the mass function of PBHs, the power spectrum of the PBH and dark matter density, and the observed black hole merger rate today.

\section{Mass function}
\label{sec:mass}

In addition to their abundance, a key (potential) observable of PBHs is their mass. We will now derive the mass function, and how it is affected by primordial non-Gaussianity. The mass function will be defined as
\begin{equation} 
\psi(m) = \frac{1}{f_{{\rm PBH}}}\frac{\mathrm{d}f_{{\rm PBH}}}{\mathrm{d}m},
\end{equation}
where $\Omega_{PBH}$ is the PBH density parameter at matter-radiation equality, and $\psi(m)$ is defined such that $\int\psi(m)\mathrm{d}m=1$, where we have followed the definition in \cite{Raidal:2018bbj} \footnote{We note that the mass function is sometimes defined to be dimensionless instead, by taking a derivative with respect to $\log m$.}. The parameter $f_{PBH}$ is the fraction of dark matter composed of PBHs, and can be calculated from $\beta$ as
\begin{equation}
f_{{\rm PBH}} = \int\limits_{M_{\rm min}}^{M_{\rm max}}\mathrm{d} (\mathrm{ln}M_{\rm H}) \left( \frac{M_{\rm eq}}{M_{\rm H}} \right)^{1/2}\beta(M_{\rm H}),
\label{eqn:omega}
\end{equation}
where $M_H$ is the horizon mass at the time of horizon-entry, $M_{\rm eq}=2.8 \times 10^{17}M_\odot$ is the approximate horizon mass at matter-radiation equality \cite{Nakama:2016gzw}. The term $\left( \frac{M_{\rm eq}}{M_{\rm H}} \right)^{1/2}$ accounts for the relative red-shift of the PBH density (which evolves like matter) relative to the dominant radiation density from the time of PBH formation until the time of matter-radiation equality. See \cite{Young:2019yug,Byrnes:2018clq} for further discussion. Assuming, for the time being, that $\zeta$ follows a Gaussian distribution, the mass function of PBHs is therefore defined completely by the primordial power spectrum $\mathcal{P}_\zeta$.


In order to derive a mass function, we will, again, make the simplifying assumption that all PBH formation occurs at a single epoch, with a Dirac-delta form for the power spectrum, as in equation \eqref{eqn:diracPower}. It is noted that such a form for the power spectrum is not physical, and the fastest growth for the power spectrum as a function of the wavevector $k$ has been shown to be $\sim k^4$ \cite{Byrnes:2018txb}, or slightly steeper in the case of a more contrived scenario \cite{Carrilho:2019oqg} (see also \cite{Bhaumik:2019tvl}), which means that PBH formation at a single scale is not physical, but rather that it occurs over a range of scales. However, our numerical calculations assuming the narrowest possible peak in the power spectrum show that the mass function would only widen by a factor $\mathcal{O}(0.5\%)$ compared to the Dirac-delta power spectrum, and so we will neglect it here in order to perform the calculations analytically. Again, the assumption of a Dirac-delta function for $\mathcal{P}_\zeta$ does not affect conclusions about the mass function, because PBH formation at each scale would be affected in the same manner. In this case, the mass function can be given as
\begin{equation}
\psi(m) = \frac{1}{\beta}\frac{\mathrm{d}\beta}{\mathrm{d}m}.
\label{eqn:psi}
\end{equation}
The mass, $m$, of a PBH is related to the amplitude of the perturbation from which it formed, by the well known critical scaling law, equation \eqref{eqn:scalingLaw}, which is inverted to give
\begin{equation}
\delta_1(m) = \frac{2}{3}\left( 2-\sqrt{4-6\delta_c-3\left( \frac{m}{\mathcal{K}M_H} \right)^{1/\gamma}}\right).
\label{eqn:delta1m}
\end{equation}
It is noteworthy here that there is a maximum value for the PBH mass which can form from perturbations of a given scale, due to the fact that equation \eqref{eqn:NLdensity} has a maximum at $\delta_R=2/3$. This gives a maximum PBH mass of
\begin{equation}
M_{max}=\mathcal{K}M_H\left(2/3-\delta_c\right)^\gamma,
\end{equation}
which gives $M_{PBH,max}=2.05M_H$ for the parameter choices considered here, see \eqref{eqn:scalingLaw}. Perturbations in $\zeta$ larger than that corresponding to the maximum $\delta_R$ are possible, but these result in perturbations of a type for which the dynamics of PBH formation are not well understood \cite{Kopp:2010sh}. In practice, this has a negligible effect on the calculation, since the abundance of such large $\zeta$ perturbations is exponentially suppressed. 

\begin{figure*}[t!]
 \centering
  \includegraphics[width=0.5\textwidth]{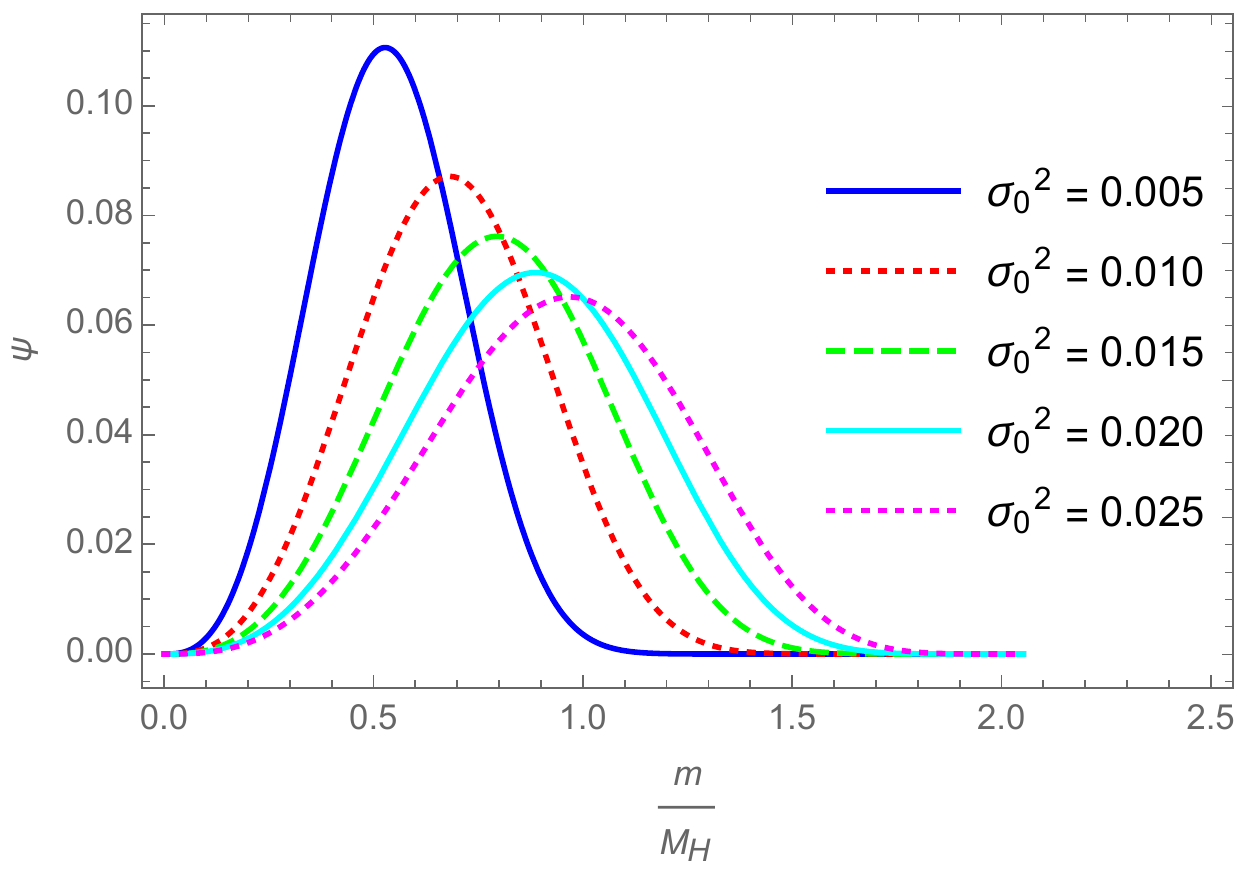} 
  \caption{ The mass function $\psi$ is shown for PBHs forming at a single time from a Dirac-delta power spectrum. The mass function generally peaks at approximately the horizon mass at the time of horizon crossing, shifting to higher masses as the variance of density perturbations, $\sigma_0^2$, increases. Note that the mass function is normalised to integrate to unity, whilst the actual abundance of PBHs is larger by many orders of magnitude for larger values of $\sigma_0^2$. } 
  \label{fig:GaussMassFunction}
\end{figure*}

Substituting equation \eqref{eqn:delta1m} into equation \eqref{eqn:psi} gives the final expression for the mass function assuming a Gaussian distribution of $\zeta$,
\begin{equation}
\psi_G(m,\sigma_0^2) = \frac{1}{4 \pi^2 \beta}\frac{m}{M_H}\left( \frac{\sigma_1}{\sigma_0} \right)^3 \left( \frac{\delta_1(m)}{\sigma_0} \right)^3\exp\left( -\frac{\delta_1^2(m)}{2\sigma_0^2} \right)\frac{\sqrt{2}\left( \frac{m}{\mathcal{K}M_H}\right)^{\frac{1}{\gamma}-1}}{\gamma\mathcal{K}M_H\sqrt{2-3\delta_c-3\left(  \frac{m}{\mathcal{K}M_H} \right)^{1/\gamma}}},
\end{equation}
where $ \frac{\sigma_1}{\sigma_0}=1$ for the Dirac-delta power spectrum considered here, and the subscript $G$ denotes that this is valid for a Gaussian $\zeta$. Figure \ref{fig:GaussMassFunction} shows the mass function for PBHs generated at a single time as a function of the horizon mass. The mass function for several values of $\sigma_0^2$ is plotted, and it can be seen that larger values for $\sigma_0^2$ mean that the mass function peaks at higher masses, but the peak is broader and lower. Note that, whilst the mass function $\psi$ is normalised to 1 when integrated over $m$, the actual abundance of PBHs is greatly suppressed for smaller values of $\sigma_0^2$ - such that more PBHs of all masses are actually produced for larger $\sigma_0^2$.

We will now turn our attention to the mass function in the presence of non-Gaussianity. As discussed in the previous section, this will be described by treating separate regions of the universe as having a Gaussian distribution, with the variance of perturbations modified by a long-wavelength perturbation, as in equation \eqref{eqn:sigmaBias}. In this case, the full mass function can be given in terms of $\psi_G$ as,
\begin{equation}
\psi\left( m, \sigma_b^2, \langle f_{\mathrm{NL}}^2\zeta_l^2 \rangle \right)=\frac{1}{\sqrt{2 \pi  \langle f_{\mathrm{NL}}^2\zeta_l^2 \rangle}}\int\limits_{-\infty}^\infty \mathrm{d}(f_{\mathrm{NL}}\zeta_l) \psi_G\left( m,\left(1+\frac{6}{5}f_{\mathrm{NL}}\zeta_l\right)^2\sigma_b^2 \right)\exp\left( -\frac{ f_{\mathrm{NL}}^2\zeta_l^2}{2\langle f_{\mathrm{NL}}^2\zeta_l^2 \rangle} \right).
\end{equation}

\begin{figure*}[t!]
 \centering
  \includegraphics[width=0.49\textwidth]{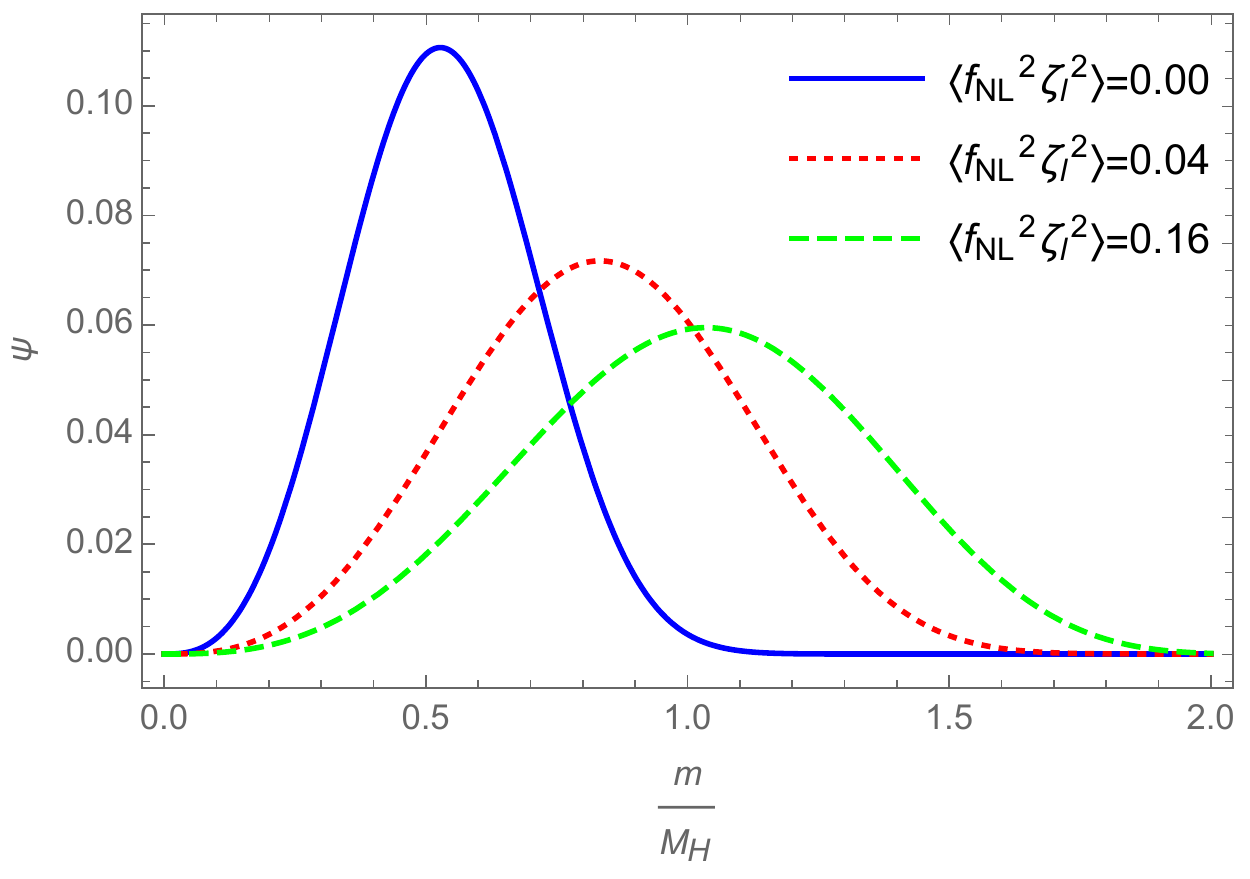} 
  \includegraphics[width=0.49\textwidth]{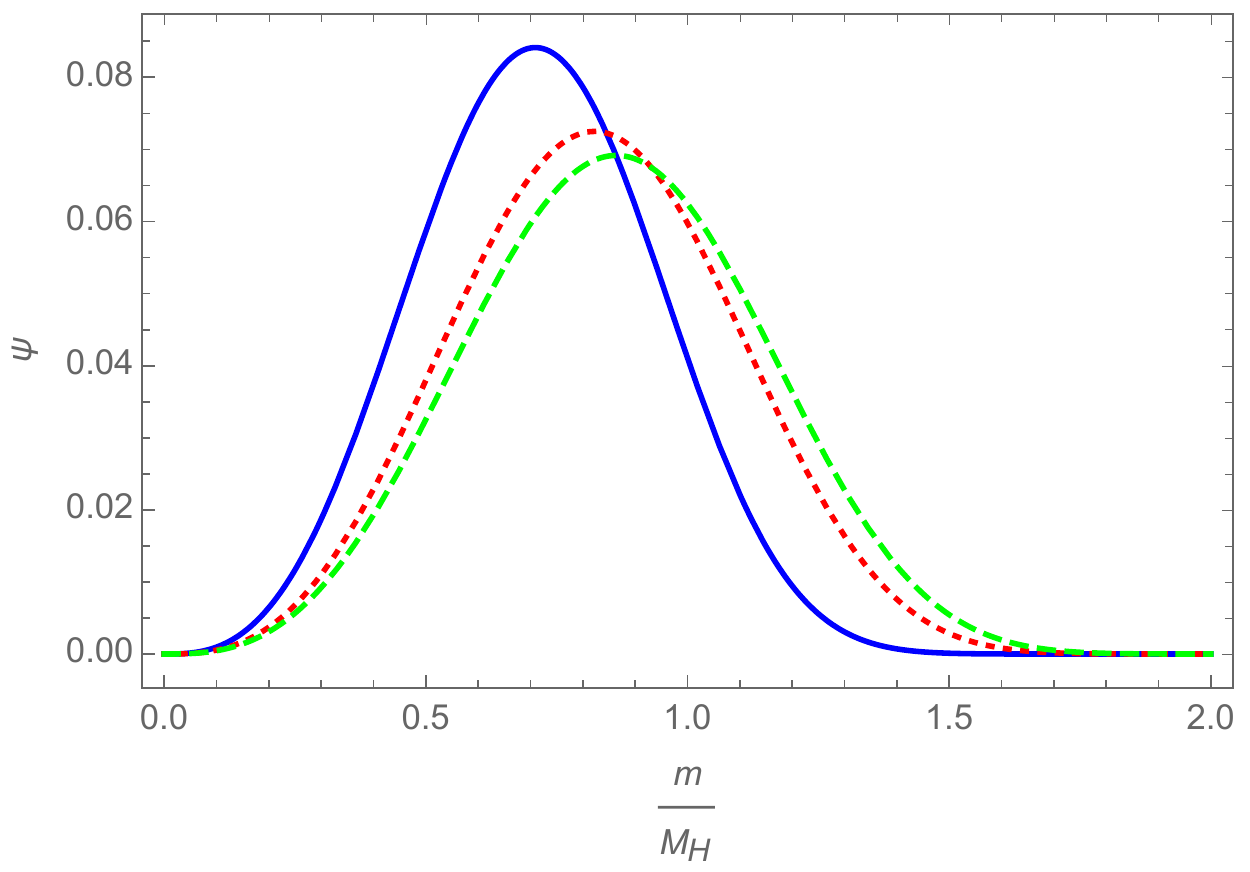} 
  \caption{ The mass function $\psi$ is shown for PBHs forming at a single time from a Dirac-delta power spectrum, this time accounting for the effect of modal coupling arising from primordial non-Gaussianity. The left plot shows the mass function for 3 strengths of modal coupling, assuming a constant background value for the variance of the density perturbations, $\sigma_b^2 = 0.005$. The right plot shows the mass function with the same strengths of modal coupling, but keeping the total abundance of PBHs fixed at $f_{\rm PBH}=0.01$.} 
  \label{fig:massFunctions}
\end{figure*}

Figure \ref{fig:massFunctions} shows the mass function for various parameter choices. The left plot shows the mass function for $\sigma_b^2=0.005$, for 3 choices of $\langle f_{\mathrm{NL}}^2\zeta_l^2 \rangle$. As $\langle f_{\mathrm{NL}}^2\zeta_l^2 \rangle$ increases, the PBHs formed in the universe become more and more clustered - such that the PBH abundance becomes quickly dominated by PBHs formed in high-$\beta$ regions. Therefore, the same behaviour for the mass spectrum is seen as for the Gaussian case: that the peak becomes lower, broader and shifted to the right as $\langle f_{\mathrm{NL}}^2\zeta_l^2 \rangle$ increases, and again the total abundance of PBHs also increases dramatically. The right plot of figure \ref{fig:massFunctions} therefore shows a similar plot, except the total abundance of PBHs is held fixed at $f_{PBH}=0.01$ for different values of $\langle f_{\mathrm{NL}}^2\zeta_l^2 \rangle$. We now see that as $\langle f_{\mathrm{NL}}^2\zeta_l^2 \rangle$ increases (which increases the PBH abundance), the background variance $\sigma_b^2$ is decreased (which decreases the PBH abundance). The change in the mass function is therefore not as pronounced, due to these factors having opposite effects, although it is still seen that the peak broadens, lowers, and shifts to the right.

This predicts a unique mass function dependent on the abundance of PBHs and the level of non-Gaussianity, but does depend on the form we have specified for the power spectrum. Given that the small-scale power spectrum is degenerate with the level of non-Gaussianity, it is not possible to use the mass function by itself (which is in principle measurable if PBHs are observed) to determine the level of non-Gaussianity or amplitude/shape of the power spectrum.

\section{The primordial black hole number density power spectrum}
\label{sec:powerSpectrum}

In this section, we will derive a simple analytic estimate for the power spectrum of the PBH density perturbations, $\mathcal{P}_{{\rm PBH}}$. To begin, we will greatly simplify the calculation of $\beta$. The dominant term in the integral in equation \eqref{eqn:beta} to calculate $\beta$ is the exponential term. Neglecting the other terms gives the standard result from the Press-Schechter calculation,
\begin{equation}
\beta = {\rm Erfc}\left( \frac{\nu_c}{\sqrt{2}} \right),
\end{equation}
where $\nu_c = \delta_{c,1}/\sigma_0$. Extending this to include the modal coupling to large-scale modes, the expression for the local PBH abundance at formation gives
\begin{equation}
\beta_{\rm local} = {\rm Erfc}\left( \frac{\nu_c}{\sqrt{2}\left(1 + \frac{6}{5}f_{\rm NL}\zeta_l \right)} \right).
\end{equation}

We will use the same definition for the PBH density perturbation $\delta_\beta$ as previously, equation \eqref{eqn:deltaPBH}, assuming that $\langle f_{\rm NL}^2\zeta_l^2 \rangle$ is small and that the total PBH abundance is not changed significantly from the background value. Expanding the resulting expression to first order in $\nu$ in the $\nu \rightarrow \infty$ limit (retaining only the leading order terms), and to first order in $f_{\rm NL}\zeta_l$ around zero gives a linear expression for $\delta_\beta$ in terms of $\zeta_l$,
\begin{equation}
\delta_\beta^{(1)} = \frac{6}{5}\nu_c^2 f_{\rm NL}\zeta_l,
\label{eqn:firstOrder}
\end{equation}
or to second-order,
\begin{equation}
\delta_\beta^{(2)} = \frac{6}{5}\nu_c^2 f_{\rm NL}\zeta_l+\frac{18}{25}\nu_c^4 f_{\rm NL}^2\zeta_l^2.
\label{eqn:secondOrder}
\end{equation}

Note that, while we have ignored the subdominant terms in equation \eqref{eqn:beta} it is possible to include them, and the full numerical result relating $\delta_\beta$ to $\zeta_l$ is shown in figure \ref{fig:deltabeta}. From this linear expression, it is simple to relate the power spectrum of $\zeta$ to the power spectrum of $\delta_\beta$,
\begin{equation}
\mathcal{P}_{\rm PBH}^{(1)}(k) = \left( \frac{6}{5}f_{\rm NL} \right)^2 \nu_c^4\, \mathcal{P}_\zeta(k),
\end{equation}
which is consistent with the expression more rigorously obtained in \cite{Suyama:2019cst}, if we have the standard picture that $ \left( \frac{6}{5}f_{\rm NL} \right)^2 = \tau_{\rm NL}$ from single-source inflation (a scenario in which any one field is responsible for generating the curvature perturbation, such as the standard curvaton or modulated reheating models) \cite{Suyama:2007bg,Byrnes:2014pja}. Including the second-order term gives
\begin{equation}
\mathcal{P}_{\rm PBH}^{(2)}(k) = \left( \frac{6}{5}f_{\rm NL} \right)^2 \nu_c^4\, \mathcal{P}_\zeta(k) + \left( \frac{6}{5}f_{\rm NL} \right)^4 \nu_c^8 \,\mathcal{P}_\zeta^2(k).
\end{equation}

\begin{figure*}[t!]
 \centering
  \includegraphics[width=0.6\textwidth]{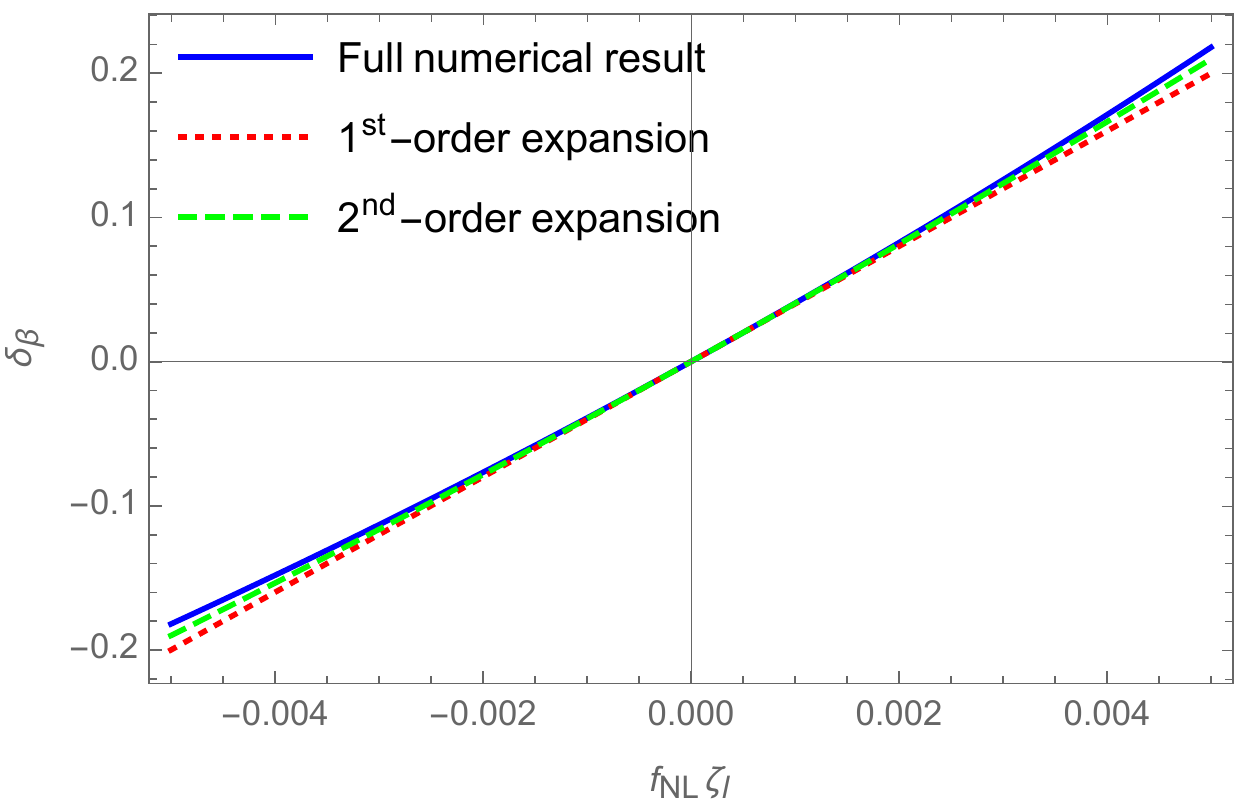} 
  \caption{ The figure shows the perturbations to the formation rate of PBHs as a function of some long-wavelength mode $\zeta_l$ multiplied by the non-Gaussianity parameter $f_{\rm NL}$. The blue line shows the full numerical result obtained with equation \eqref{eqn:deltaPBH}, whilst the dotted-red and dashed-green lines show the first and second order equations obtained by equations \eqref{eqn:firstOrder} and \eqref{eqn:secondOrder} respectively.}
  \label{fig:seriesExpansion}
\end{figure*}

Figure \ref{fig:seriesExpansion} shows a comparison of the value of $\delta_\beta$ obtained via the full numerically integrated expression given by equation \eqref{eqn:deltaPBH}, and the first- and second-order expansions obtained above. For the numerically obtained result, we have assumed a Dirac-delta form for the small-scale power spectrum as before, as in \eqref{eqn:diracPower}, with an amplitude of $\mathcal{A}_s\approx0.0938$ (such that $f_{\rm PBH}=1$ in the absence of large scale modes), giving $\nu_c = 5.77$ used in the first- and second-order expansions.

Here, contributions to the power spectrum from shot noise and adiabatic perturbations have been neglected, although they are well understood, as neither will contribute to the calculated merger rate. The shot noise has no effect because it is a consequence of the random positions of PBHs, which is already accounted for in the calculation of the expected merger rate. However, one should consider the signal arising in the power spectrum from shot noise if searching for evidence of non-Gaussianity \cite{Matsubara:2019qzv}. The adiabatic term is due to the different densities in different regions of the universe. However, at early times when PBHs form, the much larger modes under consideration here are in the super-horizon regime, and simply correspond to time-shifts in a smaller region of the universe - as described in the separate universe approach (e.g.~\cite{Wands:2000dp,Rigopoulos:2003ak}).


\section{The primordial black hole merger rate}

Over recent years, multiple observations of gravitational wave signals from the merging of binary black hole systems has been observed by LIGO \cite{LIGOScientific:2018jsj,LIGOScientific:2018mvr}, Since the paper \emph{``Did LIGO detect dark matter?''} \cite{Bird:2016dcv}, there has been an extensive amount of literature dedicated to the discussion of the formation of binary PBHs and the gravitational wave signal from merging PBHs \cite{Sasaki:2016jop,Clesse:2016vqa,Wang:2016ana,Mandic:2016lcn,Raidal:2017mfl,Nishikawa:2017chy,Kovetz:2017rvv,Raidal:2018bbj,Belotsky:2018wph,Kavanagh:2018ggo,Kohri:2018qtx,Chen:2018czv,Garriga:2019vqu}. The general consensus is that, were PBHs to make up the entirety of dark matter, the rate of observed BH merger events would be significantly higher than actually detected - thereby ruling out PBHs in the approximate mass range $1-100M_\odot$ from composing the entirety of dark matter.

In this section, we will consider the effect on the merger rate due to primordial clustering caused by PNG. We consider a scenario where PBH binaries form shortly after the formation of the PBHs themselves. At the time of formation, PBHs are coupled to the expansion of the universe, and will typically become gravitationally bound to each other when their local density becomes greater than the surrounding radiation density, normally after the time of matter-radiation equality. However, due to the randomness of the Poisson fluctuations, some PBHs will form much closer to each other than average, and will therefore decouple significantly earlier. PBH pairs may therefore start falling towards each shortly after formation, with a direct collision avoided by torque provided by gravitational forces from nearby matter perturbations and other PBHs. The pair of PBHs then forms a binary system, which will eventually merge. The merger may be observable by gravitational wave detectors.

By calculating the distribution of orbital parameters from the distribution of initial conditions leading to PBH formation, reference \cite{Raidal:2018bbj} gives an analytic expression for the merger rate of PBHs today, which reads
\begin{equation}
R = \int \mathrm{d}R_0 S,
\label{eqn:mergerRate}
\end{equation}
where $\mathrm{d}R_0$ is approximated as
\begin{equation}
\mathrm{d}R_0 = \frac{1.6\times10^6}{\mathrm{Gpc}^3\mathrm{yr}}f_{PBH}^{\frac{53}{37}}\eta^{-\frac{34}{37}} \left( \frac{m_1+m_2}{M_\odot} \right)^{-\frac{32}{37}} \left( \frac{\tau}{t_0} \right)^{-\frac{34}{37}}\psi\left(m_1\right)\psi\left(m_2\right)\mathrm{d}m_1\mathrm{d}m_2,
\end{equation}
with $\tau$ the coalescence time, $t_0=13.8\,\mathrm{Gyr}$ the current age of the universe, $m_1$ and $m_2$ the masses of the PBHs in a binary pair, and $\eta$ is the symmetric mass fraction,
\begin{equation}
\eta = \frac{m_1 m_2}{\left( m_1+m_2 \right)^2}.
\end{equation}
We will take the maximum value for the suppression factor $S$ (valid in the regime where the expected number of PBHs in a spherical volume with a radius equal to the initial separation between the PBH binary is small), given by
\begin{equation}
S = \left( \frac{5 f_{PBH}^2}{6\sigma_M^2} \right)^{\frac{21}{74}}U\left( \frac{21}{74},\frac{1}{2},\frac{5 f_{PBH}^2}{6\sigma_M^2}  \right),
\end{equation}
where $U$ is the confluent hypergeometric function, and $\sigma_M^2=\left( \Omega_M/\Omega_{DM} \right)^2 \sigma_{f}^2$ is the rescaled variance of matter density perturbations, evaluated here at the time of matter-radiation equality, which is close to the time when binaries that merge today originally formed. We will here follow previous works  \cite{Ali-Haimoud:2017rtz,Raidal:2018bbj,Eroshenko:2016hmn} and take $\sigma_{f}=0.005$. All of the factors affecting the merger rate can therefore be derived purely from the variance of density perturbations, $R = R\left(\sigma_0^2\right)$, again assuming a Dirac-delta form for the enhanced part of the power spectrum.

The expression derived by reference \cite{Raidal:2018bbj} assumed that PBHs were distributed across the universe with a uniform Poisson distribution and a constant mass function. Extending this to the non-Gaussian distribution considered here is straightforward, requiring the merger rate to be integrated over different regions of the universe in the same manner as the PBH abundance or the mass function. Calculating the local value for the merger rate also requires the calculation of the local value of the PBH DM fraction, $f_{PBH}$, and the local PBH mass function, $\psi_G$. Both of these factors depend on the local variance, $\sigma_s^2$, given in equation \eqref{eqn:sigmaBias} as a function of $f_\mathrm{NL}\zeta_l$, which follows a Gaussian distribution.

The expression for the total merger rate $R_T$ is then given as,
\begin{equation}
R_T\left( \sigma_b^2, \langle f_{\mathrm{NL}}^2\zeta_l^2 \rangle \right)=\frac{1}{\sqrt{2 \pi  \langle f_{\mathrm{NL}}^2\zeta_l^2 \rangle}}\int\limits_{-\infty}^\infty \mathrm{d}(f_{\mathrm{NL}}\zeta_l) R\left( \left(1+\frac{6}{5}f_{\mathrm{NL}}\zeta_l\right)\sigma_b^2 \right)^2\exp\left( -\frac{ f_{\mathrm{NL}}^2\zeta_l^2}{2\langle f_{\mathrm{NL}}^2\zeta_l^2 \rangle} \right).
\end{equation}

However, reference \cite{Raidal:2018bbj} also tested their analytic prediction for the merger rate with $N$-body simulations, and found that, when the PBH density becomes large, $f_{PBH}\gtrsim 0.1$, the initial binaries are likely to be disrupted by nearby PBHs which is likely to result in a reduced merger rate. As a result, equation \eqref{eqn:mergerRate} used for the merger rate cannot be trusted to be accurate for such regions. To account for this uncertainty as much as is currently possible, we have posited several different models. We have introduced a ``hard'' and ``soft'' cut-off in the integral above, where the merger rate is set to zero or to the value corresponding to $f_{PBH}=0.1$ in regions where $f_{PBH} > 0.1$, respectively.

\begin{figure*}[t!]
 \centering
  \includegraphics[width=0.6\textwidth]{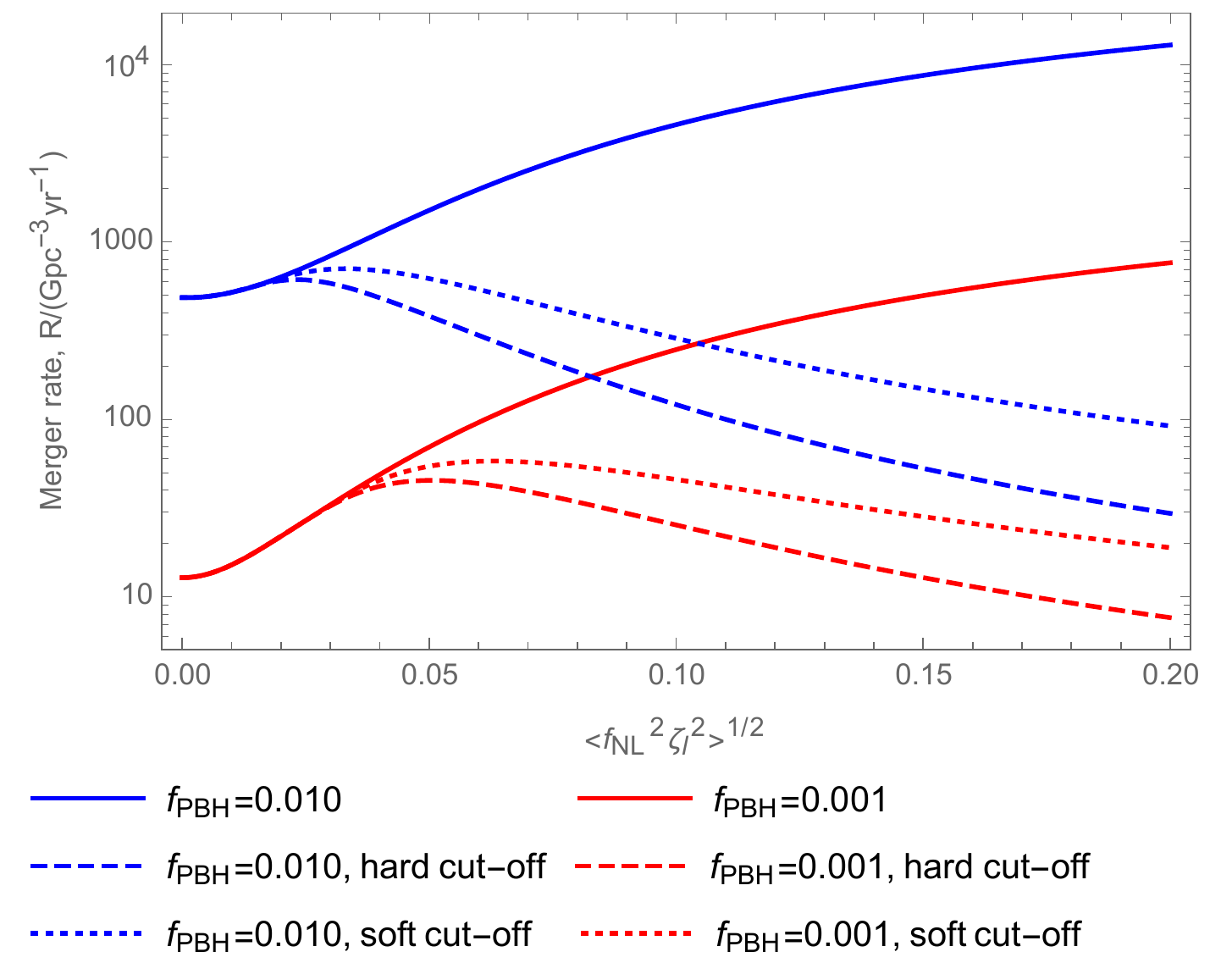} 
  \caption{ The effect of local-type non-Gaussianity is plotted against the predicted merger rate for PBHs. The merger rate is plotted for 2 different total abundances of PBHs, $f_{\rm PBH}=0.01$ and $f_{\rm PBH}=0.001$ in blue and red respectively. For small amounts of clustering, the merger rate is expected to increase, but there is significant uncertainty when the clustering becomes very large - due to the uncertainty in the evolution of binary systems in PBH-dense regions.} 
  \label{fig:mergerRate}
\end{figure*}

The merger rate today will then depend on 2 factors: the abundance of PBHs, and the level of non-Gaussianity, parameterised by $f_{PBH}$ and $\langle f_{\mathrm{NL} }^2\zeta_l^2\rangle$ respectively. Figure \ref{fig:mergerRate} shows the predicted merger rate today as a function of $\langle f_{\mathrm{NL}}^2\zeta_l^2\rangle$ for $f_{PBH}=0.01$ and $0.001$. Where $\langle f_{\mathrm{NL}}^2\zeta_l^2\rangle=0$, our expression is reduced to that given by \cite{Raidal:2018bbj}. It can be seen that as the level of non-Gaussianity increases, the clustering also increases, which leads to a rise in the predicted merger rate in all cases. However, as the level of non-Gaussianity continues to rise, the PBH abundance in the universe starts being dominated by regions where the local PBH density is very high - which means equation \eqref{eqn:mergerRate} can no longer be trusted. In this case, the merger rate may 
start dropping - as shown by the dashed- and dotted-lines, representing hard and soft cut-offs respectively. Above the point at which the lines diverge, there is still a great deal of uncertainty in what the merger rate may be. The dashed and solid lines can be considered as lower and upper bounds respectively, with a ``more realistic'' estimate given by the dotted lines.

\subsection{Smallest value of $f_{{\rm PBH}}$ which might produce the observed merger rate}

In this section, we will provide a simple order-of-magnitude estimate for the lower limit of PBH abundance which may give rise to the observed BH-BH mergers observed by LIGO. Whilst there are significant uncertainties arising in the calculation from the current uncertainty in the merger rate when the local PBH abundance becomes large, in this section we will neglect this in order to give a rough estimate for the lower bound. 

We will assume a simple picture where PBHs form in extremely high abundance in certain regions (hereafter referred to high PBH density (HPD) regions), and in negligible amounts elsewhere\footnote{A recent paper \cite{Ding:2019tjk} used a similar model, concluding that initial clustering increases the detectability of PBHs via gravitational waves from merging events.}. Such a scenario may be realised within the framework of our model if the variance of $f_{\rm NL}\zeta_l$ is extremely large, leading to a small number regions with a very large small-scale power spectrum with high PBH abundance, and a large number of regions with small power spectrum and low PBH abundance (although the full calculation described above breaks down when PBHs can no longer be considered rare events).

In order to provide the required number of PBH mergers, we will take that the minimum merger rate to be $10\,{\rm Gpc}^{-3}{\rm yr}^{-1}$, that PBHs form when the horizon mass is equal to $20 \,M_\odot$, and that the maximum formation rate in any HPD region is $\beta=0.1$. In such HPD regions, the PBH energy density would be $\mathcal{O}(10^7)$ times greater than the dark matter density in the universe, $f_{\rm PBH}\sim 10^7$. The method described in section \ref{sec:mass} breaks down when PBHs are not rare events, and so we will assume a lognormal mass function given by, 
\begin{equation}
\psi(m) = \frac{1}{\sqrt{2\pi}\sigma_m m_c}\exp \left( -\frac{-\log^2(m/m_c)}{2\sigma_m^2} \right),
\end{equation}
where we take values $m_c=20\,M_\odot$ and $\sigma_m=0.1$ (the result is not especially sensitive to the exact values). The merger rate predicted by equation \eqref{eqn:mergerRate} is then $R = \mathcal{O}(10^{15})$. In order to produce the minimum merger rate of $10\,{\rm Gpc}^{-3}{\rm yr}^{-1}$, such regions must therefore occupy approximately $10/10^{15}=10^{-14}$ of the entire volume of the universe. Therefore, the minimum fraction of dark matter composed of PBHs in order to be responsible for the BH-BH merger events observed by LIGO is
\begin{equation}
f_{\rm PBH}\approx10^7 \times 10^{-14} = 10^{-7}. 
\end{equation}
We stress that this number is intended to be a crude order--of--magnitude estimate only, and assumes that the analytic prediction for the merger rate holds for (locally) very large PBH abundances. In reality, the lower bound is likely to be significantly higher.

\section{Summary}
\label{sec:summary}

We have studied the way in which primordial non-Gaussianity may affect key observables related to the production of PBHs in the early universe. Unlike for a purely Gaussian distribution, primordial non-Gaussianity arising from inflation, especially local-type, is expected to result in significant modal coupling. We have studied the effect that such modal coupling would have on the abundance of PBHs, the mass function, the power spectrum of PBH number density perturbations on larger scales at the time of formation, and the merger rate of PBHs observable today from binary PBH systems. 

The effect of modal coupling is found to always increase the abundance of PBHs, regardless of the sign of $f_{\rm NL}$. This is due to the exponential dependance of the PBH abundance on the amplitude of the power spectrum. In the case of positive (negative) $f_{\rm NL}$, PBH formation is enhanced in regions of positive (negative) $\zeta_{l}$, and vice-versa. This leads to a significant variation in the formation rate of PBHs in different regions of the universe, $\delta_\beta$. In section \ref{sec:powerSpectrum}, we calculated the power spectrum $\mathcal{P}_{\rm PBH}$ to second order in terms of the power spectrum of the curvature perturbation $\mathcal{P}_\zeta$. A possible future observation of such perturbations could therefore provide insights into the early universe.

We have discussed the mass function of PBHs which form, assuming a very narrow peak in the power spectrum - in our case, a Dirac-delta peak. The mass function peaks at approximately the horizon mass at the time perturbations enter the horizon. The location of the peak shifts to higher masses when the amplitude of the power spectrum is larger (or in regions where the local amplitude is higher) - corresponding to an increase in the formation rate of PBHs. Therefore, in addition to the variance in the abundance of PBHs in different regions of the universe, variations in the mass function of PBHs may, in the future, provide information about primordial physics.

The effect of modal coupling from non-Gaussianity on the possible observed GW signal today from merging binary PBH systems has also been considered. In the case that the modal coupling is relatively small, $\langle f_{\rm NL}^2\zeta_l^2 \rangle \lesssim 4\times 10^{-4}$, the effect of modal coupling can be safely stated to increase the predicted merger rate. This means that much smaller abundances of PBHs than previously thought could still produce a high enough present day merger rate.
 We have estimated that the minimum amount of dark matter composed of PBHs which could still be responsible the merger events observed by LIGO is $f_{\rm PBH}\sim10^{-7}$. A lower PBH abundance is therefore excluded from producing the observed GW signals.

However, the formation rate of binary systems in regions of high PBH abundance (i.e.~where the PBH density is greater than $10\%$ of the background dark matter density) is currently not well understood - and this means that there is significant uncertainty in our calculation of the merger rate when the initial clustering becomes too large. It may even have the effect of reducing the merger rate sufficiently such that the merger rate constraints on $f_{\rm PBH}$ may be weakened if there is significant PBH clustering. We note that a paper which appeared while we were completing this work \cite{Vaskonen:2019jpv} have estimate the weakest possible constraint from the merger rate by considering the merger rate of PBHs from binary pairs which have been disrupted. PBH abundances previously thought ruled out by the lack of observed merger events may not be ruled out.

Throughout this paper, we have ignored the effect of non-Gaussianity on PBH scales - although the results presented would be qualitatively unchanged by its inclusion. Full consideration of this will require consideration of the the shape of the small-scale power spectrum, typical profile shapes and the shape of the bispectrum, amongst other factors, and goes beyond the scope of this paper. However, much work has been completed on this topic in the past \cite{Byrnes:2012yx,Shandera:2012ke,Young:2013oia,Young:2014oea,Young:2015cyn,Franciolini:2018vbk,Yoo:2019pma,Atal:2019cdz,Kalaja:2019uju}, and a more complete analysis using the methods presented in this paper would yield qualitatively similar results. We have also only considered an expansion to second order in local-type non-Gaussianity, although it has been shown that higher-order terms can also be important \cite{Young:2013oia}.


\section*{Acknowledgements}
SY is funded by a Humboldt Research Fellowship for Postdoctoral Researchers. CB is supported by a Royal Society University Research Fellowship.


\bibliographystyle{JHEP} 
\bibliography{bibfile}

\end{document}